\documentclass[
  aps,
  prb,
  twoside,
  twocolumn,
  preprintnumbers,
  floatfix,
  showpacs,
  superscriptaddress,
]{revtex4-1}

\usepackage{graphicx}
\usepackage{amsmath}
\usepackage{amssymb}
\usepackage{times}
\usepackage{subfigure}
\usepackage{color}

\newcommand{\A}{\text{A}}
\newcommand{\B}{\text{B}}
\newcommand{\VCSGC}{\text{V}}
\newcommand{\SGC}{\text{S}}
\newcommand{\C}{\text{C}}
\newcommand{\CPU}{\text{CPU}}

\newcommand{\eV}{\text{eV}}
\newcommand{\etal}{\textit{et al.}}

\renewcommand{\vec}[1]{\ensuremath\boldsymbol{#1}}
\renewcommand{\emph}[1]{\textit{#1}}

\providecommand{\sect}[1]{Sect.~\ref{#1}}
\providecommand{\eq}[1]{(\ref{#1})}
\providecommand{\Eq}[1]{Eq.~\eq{#1}}

\providecommand{\fig}[1]{Fig.~\ref{#1}}

\begin{document}

\title{
  A scalable parallel Monte Carlo algorithm \\
  for atomistic simulations of precipitation in alloys
}

\author{Babak Sadigh}
\email{sadigh1@llnl.gov}
\affiliation{
  Lawrence Livermore National Laboratory,
  Condensed Matter and Materials Division,
  Livermore, California, USA
}
\author{Paul Erhart}
\email{erhart@chalmers.se}
\affiliation{
  Lawrence Livermore National Laboratory,
  Condensed Matter and Materials Division,
  Livermore, California, USA
}
\affiliation{
  Chalmers University of Technology,
  Department of Applied Physics,
  Gothenburg, Sweden
}
\author{Alexander Stukowski}
\affiliation{
  Lawrence Livermore National Laboratory,
  Condensed Matter and Materials Division,
  Livermore, California, USA
}
\author{Alfredo Caro}
\author{Enrique Martinez}
\affiliation{
  Lawrence Livermore National Laboratory,
  Condensed Matter and Materials Division,
  Livermore, California, USA
}
\affiliation{
  Los Alamos National Laboratory,
  Los Alamos, New Mexico, USA
}
\author{Luis Zepeda-Ruiz}
\affiliation{
  Lawrence Livermore National Laboratory,
  Condensed Matter and Materials Division,
  Livermore, California, USA
}

\begin{abstract}
We present an extension of the semi-grandcanonical (SGC) ensemble that we refer to as the variance-constrained semi-grandcanonical (VC-SGC) ensemble. It allows for transmutation Monte Carlo simulations of multicomponent systems in multiphase regions of the phase diagram and lends itself to scalable simulations on massively parallel platforms. By combining transmutation moves with molecular dynamics steps structural relaxations and thermal vibrations in realistic alloys can be taken into account. In this way, we construct a robust and efficient simulation technique that is ideally suited for large-scale simulations of precipitation in multicomponent systems in the presence of structural disorder. To illustrate the algorithm introduced in this work, we study the precipitation of Cu in nanocrystalline Fe.
\end{abstract}

\date{\today}

\pacs{
  02.70.Tt  
  05.10.Ln  
  81.07.Bc  
  81.30.Mh  
}

\maketitle


\section{Introduction}

The interplay between chemistry and structure is of paramount importance in materials science. This applies in particular to alloys where chemical ordering and precipitation in the presence of surfaces, grain boundaries, dislocations and other structural features lead to complex behavior. Some examples of practical importance include Al-Cu alloys, Ni-Co superalloys as well as steels, the properties of which vary over a wide range depending on composition and microstructure. Understanding and eventually controlling these effects is a prerequisite for designing and improving materials. In principle, modeling and simulation are ideally suited to complement and guide experimental efforts, especially as dimensions shrink and chemical complexity increases.

The objective of the present work is to develop an algorithm that enables us to model the equilibrium properties of phase segregated multicomponent systems containing millions of particles while taking into account chemical degrees of freedom, structural relaxations as well as thermal vibrations. For such an algorithm to be useful on current computing platforms, it must lend itself to efficient parallelization. This is difficult to achieve for Monte Carlo (MC) algorithms that are based on the canonical ensemble. \cite{FreSmi01} Simulations within the semi-grandcanonical (SGC) ensemble on the other hand are easily parallelized but cannot be used to study precipitation and interface formation. The objective of the present work is to develop a MC technique that both can handle multiphase systems and be parallelized easily and efficiently. Note that the parallel algorithm discussed in this paper is suitable for short-range interatomic potentials as described e.g., by embedded-atom method, \cite{DawBas84} bond-order, \cite{Ter86} or Stillinger-Weber \cite{StiWeb85} type potentials. 

The paper is organized as follows. In \sect{sect:chemical_mixing}, we discuss how to model chemical mixing and phase segregation on the atomic scale. The most common approach is to sample the chemical configuration space using transmutational MC methods, which require as key ingredient an appropriate statistical ensemble. Following a discussion of the advantages and shortcomings of existing ensembles with respect to the present application, we introduce the variance-constrained semi-grandcanonical (VC-SGC) ensemble, which can be viewed as a generalization of the extended Gaussian ensemble technique to multicomponent systems, \cite{Het87, JohPlaViv03} and formulate a simple serial VC-SGC-MC algorithm. In \sect{sect:parallelization}, we address the question how the MC methods introduced in \sect{sect:chemical_mixing} can be adapted for simulations of systems containing millions of particles. To this end, we derive transition matrices and their efficient decomposition. In \sect{sect:applications}, we finally discuss the simultaneous and efficient sampling of chemical, structural and vibrational degrees of freedom, and consider the precipitation of Cu in nanocrystalline Fe as an illustrative example.

The algorithms developed in this work have been implemented in the massively parallel molecular dynamics code \textsc{lammps} \cite{Pli95} and the source code is available from the authors.


\section{Modeling chemical mixing and precipitation}
\label{sect:chemical_mixing}

On the atomic scale, chemical mixing in alloys is most commonly studied using MC simulations within either the semi-grandcanonical (SGC) or the canonical ensemble. Therefore, we first discuss in some detail these two ensembles before deriving the variance-constrained semi-grandcanonical ensemble (VC-SGC), which merges the advantages of the canonical and semi-grandcanonical ensembles. In the following, we use the subscripts C, S, and V to indicate quantities that are connected to the canonical, SGC and VC-SGC ensembles, respectively.  For the sake of simplicity, we limit our discussion to binary alloys. The generalization to systems containing an arbitrary number of species is straightforward.

Consider a system of $N$ particles confined in a box of volume $V$, where each particle carries a spin of value 0 or 1. A configuration of this system can be denoted $(\vec{x}^{3N},\sigma^N)$, where $\vec{x}^{3N}$ is a $3N$-dimensional vector describing the positions of every particle, and $\sigma^N$ is an $N$-dimensional spin vector. The number of spin 1 particles is $n=\sum_{i=1}^N \sigma_i$, and their concentration $c=n/N$. We denote the energy of a configuration by $U(\vec{x}^{3N},\sigma^N)$.


\subsection{The canonical ensemble}
\label{sect:canonical}

The canonical ensemble describes the thermodynamics of systems that are chemically isolated, i.e. the number of members of each species is kept constant. The partition function for the canonical ensemble at temperature $T$ for the binary system defined above is 
\begin{align}
  \mathcal{Z}_{\C}\left(c,\mathcal{N}\right)
  &= \Lambda_1^{-3(N-n)}\Lambda_2^{-3n} \frac{1}{n!(N-n)!}
  \nonumber
  \\
  &\quad\quad
  \int \exp\left[-\beta U\left(\vec{x}^{3N},\sigma^{N}\right)\right]~d^{3N}\vec{x},
  \label{eq:Zc0}
\end{align}
where $\beta=1/k_B T$, $\Lambda_i=\sqrt{h^2/2\pi m_ikT}$ is the thermal de Broglie wavelength for component $i$, and $\mathcal{N}=\{N,V,T\}$  is the set of independent thermodynamic variables.\cite{FreSmi01} Monte Carlo simulations in this ensemble sample the probability distribution
\begin{align}
  \pi_{\C}\left(\vec{x}^{3N},\sigma^N;c,\mathcal{N}\right)  
  \propto \exp\left[ -\beta U\left(\vec{x}^{3N},\sigma^N\right) \right]. 
\end{align}
Efficient sampling of the above distribution involves two kinds of trial moves: ({\em i}) particle displacements $\vec{x}^{3N}\rightarrow \vec{x}^{3N}_t$, and ({\em ii}) compositional changes $\sigma^{N}\rightarrow \sigma_t^{N}$ that keep the concentration fixed. In practice, in trial move ({\em i}) a particle is selected at random and assigned a random displacement, while for trial move ({\em ii}) two particles with unlike spins are selected at random and their spins are exchanged. These trial moves are accepted with probability
\begin{align}
  \mathcal{A}_{\C} &=
  \min\left\{1,
  \exp\left[ - \beta \Delta U \right]
  \right\} ,
  \label{eq:trans_canonical}
  \\
  \Delta U &= U(\vec{x}_t^{3N},\sigma_t^N) - U(\vec{x}^{3N},\sigma^N).
  \label{eq:dU}
\end{align}
This acceptance probability is designed to satisfy detailed balance. Approach to equilibrium can be accelerated substantially if trial moves ({\em i}) are biased along the force vector $-\vec{\nabla}U(\vec{x}^{3N},\sigma^N)$. This is achieved most easily via a hybrid technique where particle positions $\vec{x}^{3N}$ are sampled via molecular dynamics (MD) while spin degrees of freedom are sampled using the spin exchange (transmutation) MC moves described above.


\subsection{The semi-grandcanonical ensemble}
\label{sect:sgc}

The SGC ensemble describes the thermodynamics of a system in contact with an infinite reservoir at constant temperature and chemical potential for each species. This ensemble corresponds to a set of configurations with varying compositions, but with their ensemble average constrained by the reservoir. The equilibrium probability distribution of the SGC ensemble for the binary system defined above thus becomes
\begin{align}
  \pi_{\SGC}(\vec{x}^{3N},\sigma^N; \Delta\mu,\mathcal{N})
  &\propto \exp\left[
    -\beta ( U(\vec{x}^{3N},\sigma^N) + \Delta\mu N\hat{c}(\sigma^N) ) \right]
  \nonumber
  \\
  \hat{c}(\sigma^N)
  &= \frac{1}{N}\sum_{i=1}^N\sigma_i,
  \label{eq:pi_sgc_orig}
\end{align}
where $\Delta\mu$ is a Lagrange multiplier that constrains the average concentration. The partition function can be expressed in terms of the canonical one via
\begin{align}
  \mathcal{Z}_{\SGC}(\Delta\mu,\mathcal{N})
  &= 
  \int_0^1 \mathcal{Z}_{\C}(c,\mathcal{N})  \exp\left[-\beta\Delta\mu Nc\right]~dc.
\end{align}

The SGC ensemble can be sampled using a Monte Carlo algorithm, where trial moves $\sigma^N\rightarrow\sigma_t^N$ are made by ({\em i}) selecting a particle at random, ({\em ii}) flipping its spin, ({\em iii}) computing the change in energy $\Delta U$, and concentration $\Delta c$. Trial moves are accepted with probability 
\begin{align}
  \mathcal{A}_{\SGC} &=
  \min \left\{ 1,
  \exp\left[ -\beta (\Delta U + \Delta\mu N\Delta c) \right]
  \right\},
  \label{eq:trans_sgc}
\end{align}
which is designed to satisfy detailed balance. 

\begin{figure}
  \centering
\includegraphics[scale=0.65]{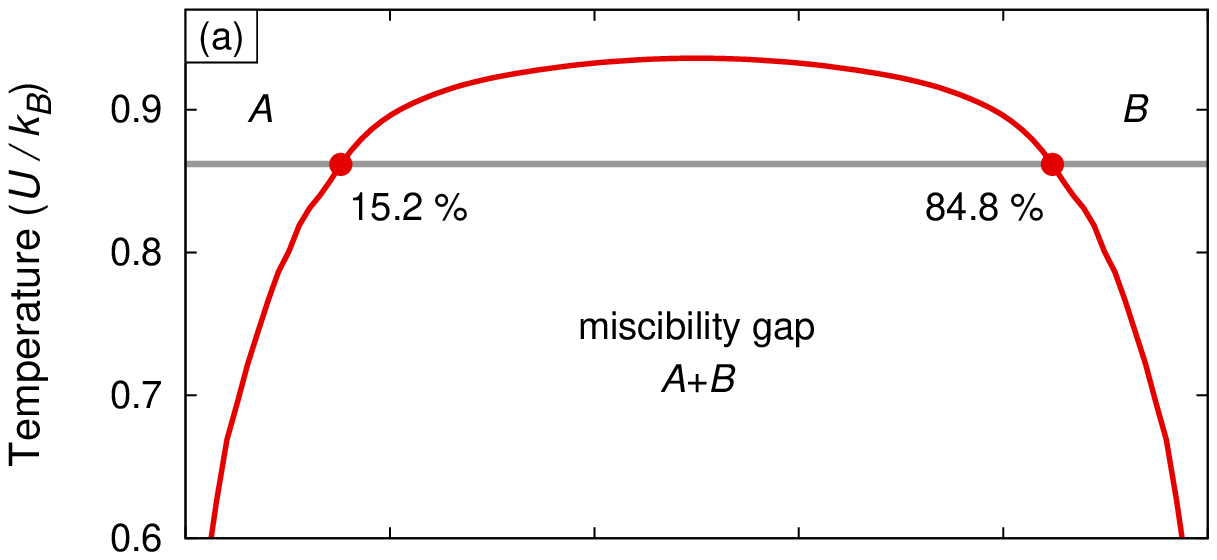}
\includegraphics[scale=0.65]{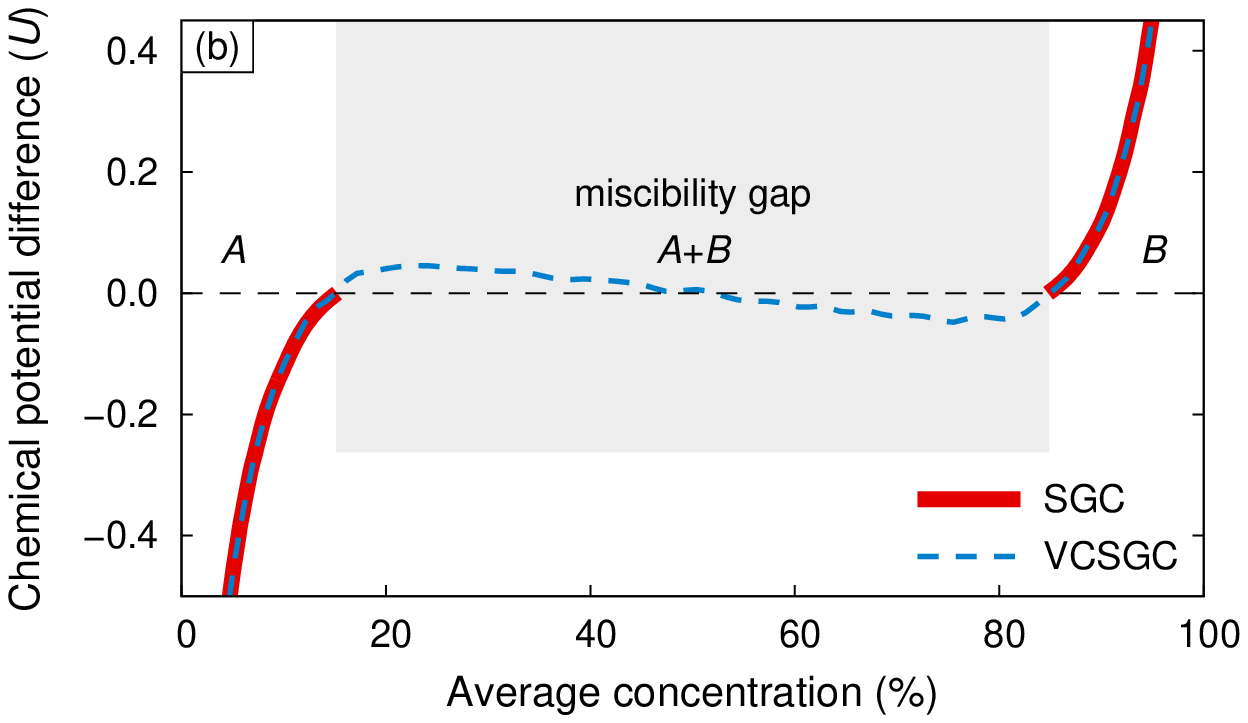}
  \caption{
    (a) Phase diagram for the Ising-type model system described in the text. The horizontal bar marks the temperature of $0.8617\,U/k_B$ at which all the simulations described in this paper have been carried out. The closed circles indicate the solubility limits at this temperature.
    (b) The chemical driving force $\Delta\mu$ as a function of concentration as obtained from a series of simulations in the semi-grandcanonical (SGC, solid line) and variance-constrained semi-grandcanonical (VC-SGC, dashed line) ensembles, respectively.
  }
  \label{fig:phasediagram}
  \label{fig:df1_sgc}
  \label{fig:df1_vcsgc}
\end{figure}

The acceptance probability expression above has important physical significance. It shows that in the SGC ensemble the force associated with a change in the chemical configuration does not solely originate from the potential energy function $\Delta U$, but also from the term $\Delta\mu N\Delta c$. In particular, for any change in concentration, a constant external chemical driving force $\Delta\mu N$ is added to the usual interatomic forces in order to drive the equilibrium concentration to the desired value. In physical experiments $\Delta\mu$ corresponds to the chemical potential difference between the two species. In practice it alters the acceptance probability \eq{eq:trans_sgc} for trial moves that lead to a concentration change. It is important to note that in this way only single-phase equilibria can be established. This means that e.g., for immiscible systems such as the one shown in \fig{fig:phasediagram}(a), concentrations inside the miscibility gap cannot be stabilized. This limitation results from the functional dependence between the chemical potential difference $\Delta\mu$ and average concentration $\left<\hat{c}\right>_{\SGC}$ not being one-to-one in the multiphase regions of the phase diagram. 

To illustrate this point, let us consider an Ising-type Hamiltonian
\begin{align}
  \mathcal{H}
  &= \frac{1}{2} \sum_{i\in\A,j\in\A} \epsilon_{\A\A}(r_{ij})
  \nonumber \\
  &\quad\quad
  +  \frac{1}{2} \sum_{i\in\A,j\in\B} \epsilon_{\A\B}(r_{ij})
  +  \frac{1}{2} \sum_{i\in\B,j\in\B} \epsilon_{\B\B}(r_{ij})
\end{align}
where $r_{ij}$ denotes the neighbor shell of site $i$ in which site $j$ is located. We use a body-centered cubic (BCC) lattice with interactions up to the second neighbor shell and $\epsilon_{AA}(1)=\epsilon_{BB}(1)=-10\,U$, $\epsilon_{AB}(1)=-9.7\,U$, and $\epsilon_{AA}(2)=\epsilon_{BB}(2)=\epsilon_{AB}(2)=-2\,U$. The phase diagram for this model system can be calculated analytically and is shown in \fig{fig:phasediagram}(a). We carried out a series of simulations using the SGC-MC method for a system containing 2000 sites at a temperature of $0.8617\,U/k_B$, starting from a solid solution at 50\%. The dependence of $\Delta\mu$ on $\left<\hat{c}\right>_{\SGC}$ determined in this way is depicted by the solid red line in \fig{fig:df1_sgc}(b). Note the discontinuity in the $\Delta\mu$--$\left<\hat{c}\right>_{\SGC}$ plot, which occurs in the region of the binary phase diagram where the miscibility gap is located. This demonstrates that the SGC-MC method is not suitable for studying phase segregation.


\subsection{The variance-constrained semi-grandcanonical ensemble}
\label{sect:vcsgc}

\begin{figure}
  \centering
\includegraphics[scale=0.65]{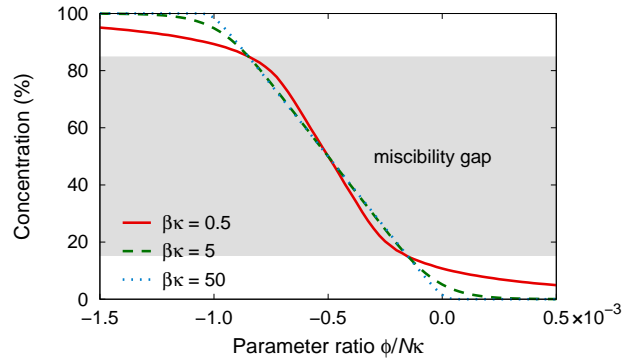}
  \caption{
    Dependence of global concentration on the parameter ratio $\phi/N\kappa$ obtained from VC-SGC-MC simulations. All simulations were carried out at a temperature of $0.8617\,U/k_B$ for the model system described in \sect{sect:sgc}.
  }
  \label{fig:vcsgc_conc_mu_eps}
\end{figure}

To simulate systems in multiphase regions of phase diagram, where precipitation occurs, we modify the SGC ensemble by adding a constraint that fixes the ensemble-averaged squared concentration $\left<\hat{c}^2\right>$. This limits concentration fluctuations and thus, when inside the miscibility gap, prevents the concentration to fluctuate to the phase boundaries. We refer to this approach as the variance-constrained semi-grandcanonical (VC-SGC) ensemble, which can be categorized as an extended Gaussian ensemble. Such ensembles describe the thermodynamics of systems in contact with finite reservoirs. \cite{JohPlaViv03} We will show below that the VC-SGC ensemble is ideal for studying equilibrium properties of multiphase systems and that it is quite straightforward to devise Monte Carlo algorithms that sample this ensemble.

\begin{figure}
  \centering
\includegraphics[scale=0.65]{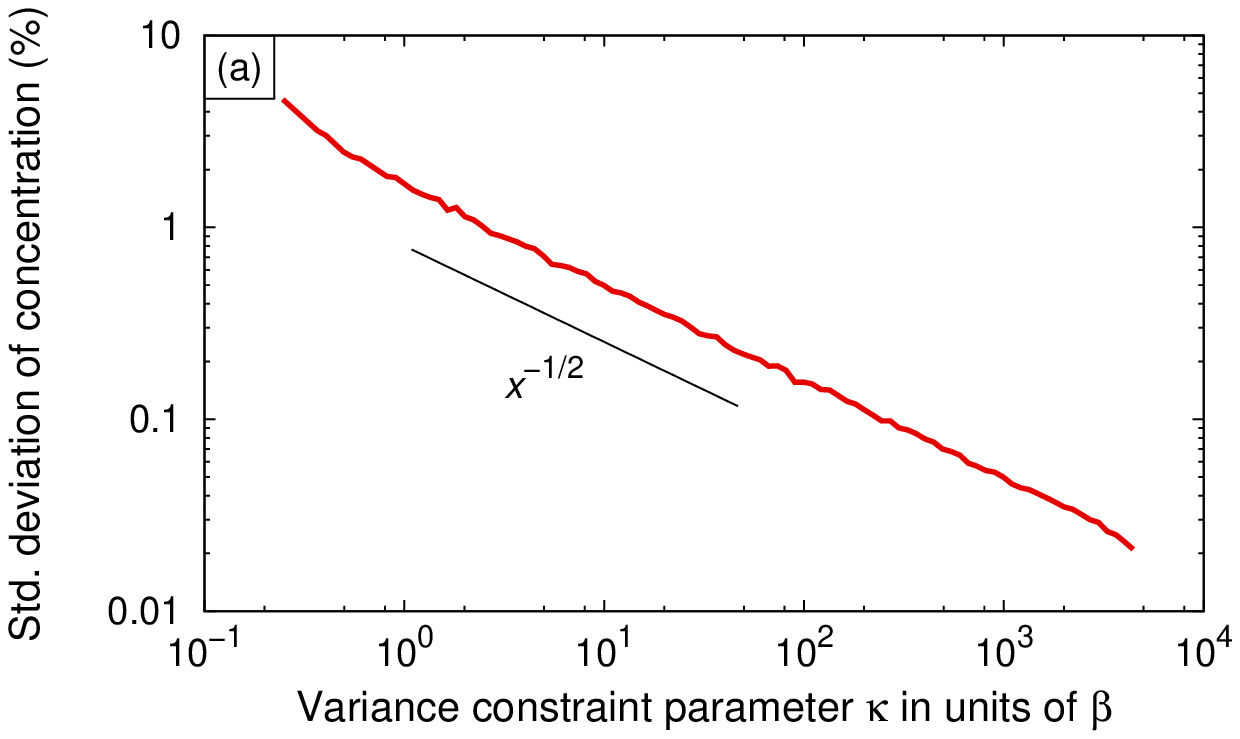}
\includegraphics[scale=0.65]{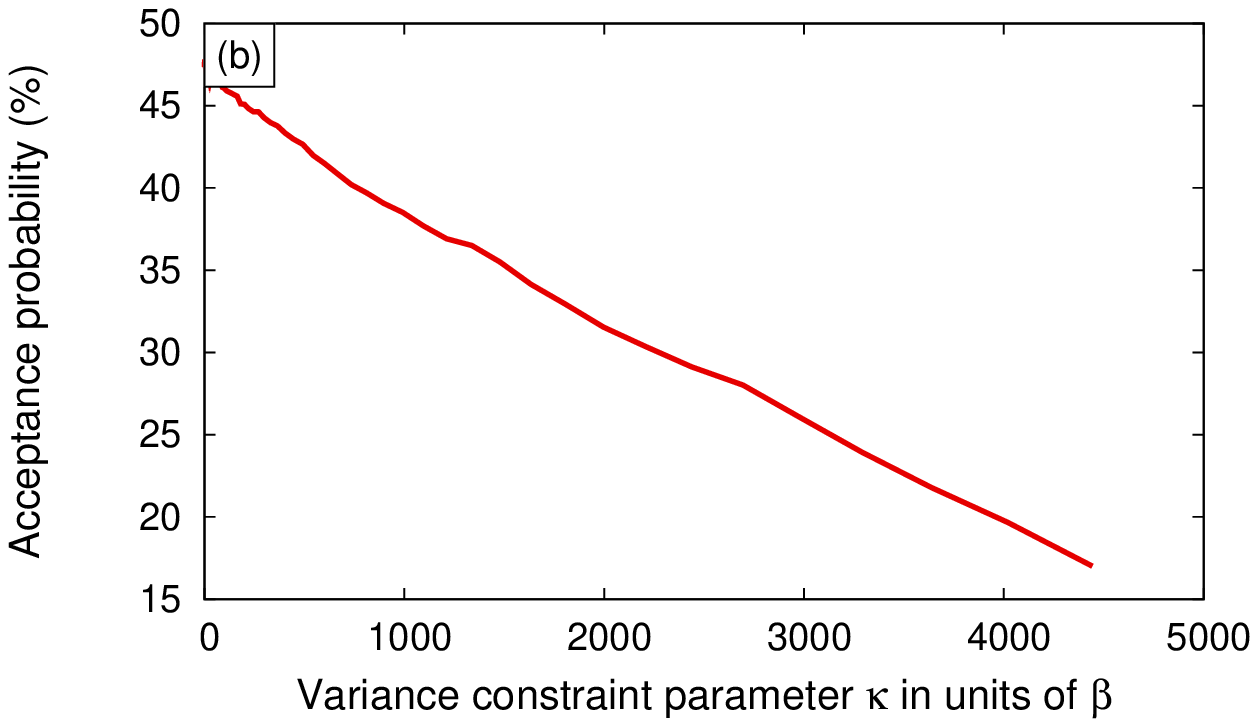}
  \caption{
    Dependence of (a) standard deviation of concentration and (b) acceptance probability on the variance constraint parameter
    $\kappa$.
  }
  \label{fig:vcsgc_stddev_eps}
  \label{fig:vcsgc_accprob_eps}
\end{figure}

In contrast to the SGC ensemble that is characterized by an infinite reservoir with constant chemical potential $\Delta\mu$, the reservoir of the VC-SGC ensemble is controlled by two independent parameters $\phi$ and $\kappa$. The statistical mechanical origin of these parameters is laid out in detail in the appendix. There it is shown that $\phi$ and $\kappa$ are Lagrange multipliers associated with constraints on the first and the second moments of the concentration, respectively. The most probable distribution subject to these constraints is then derived to be (see Eq.~\ref{eq:vprob})
\begin{align}
  \pi_{\VCSGC}\left(\vec{x}^{3N},\sigma^N;\phi,\kappa,\mathcal{N}\right)
 \propto
 \exp\left[ -\beta U(\vec{x}^{3N},\sigma^N) \right]
 \label{eq:pi_vcsgc}
 \\
 \quad \times \exp\left[ - \beta N\hat{c}(\sigma^N) \left(\phi + 
   \kappa N \hat{c}(\sigma^N)\right) \right] \nonumber.
\end{align}
We can thus express the partition function of the VC-SGC ensemble in terms of the canonical one as
\begin{align}
  \mathcal{Z}_{\VCSGC}(\phi,\kappa,\mathcal{N})  = 
  \int_0^1 \mathcal{Z}_{\C}(c,\mathcal{N})  \exp\left[-\beta N c(\phi+\kappa Nc)\right]~dc.
\end{align}

The VC-SGC ensemble can be considered a generalization of both the SGC and the canonical ensembles. The former is obtained trivially by letting $\kappa\rightarrow 0$. In order to obtain the canonical ensemble, we complete the square in \Eq{eq:pi_vcsgc} and rewrite the VC-SGC probability distribution as
\begin{align}
  \pi_{\VCSGC}\left(\vec{x}^{3N},\sigma^N;\phi,\kappa,\mathcal{N}\right)
  & \propto
  \exp\left[ -\beta U(\vec{x}^{3N},\sigma^N) \right]
  \\
  & \times \exp\left[ - \beta \kappa \left( N\hat{c}(\sigma^N)
    +\frac{\phi}{2\kappa}\right)^2 \right]
  \nonumber.
\end{align}
The canonical ensemble is recovered when $\kappa\rightarrow\infty$ and $\phi=-2\kappa Nc$. This can be seen by rewriting the canonical partition function as
\begin{align}
  \mathcal{Z}_{\C}\left(c,\mathcal{N}\right) = 
  \int_0^1 \mathcal{Z}_{\C}(c',\mathcal{N})  \delta\left(c-c'\right)~dc'.
\end{align}
Hence the VC-SGC ensemble may be obtained by generalizing the delta function that fixes the concentration in the canonical ensemble to a Gaussian with tunable width determined by the parameter $\kappa$. Now in multiphase regions of phase diagrams, where the SGC ensemble is not stable, a VC-SGC ensemble can be devised by judiciously choosing the two parameters $\phi$ and $\kappa$ that combine both advantages of the SGC and the canonical ensembles. Traditionally the canonical ensemble has been used to study precipitation inside the miscibility gap. Our objective with this paper is to show that the same physics can be studied much more efficiently in the VC-SGC ensemble, especially when parallel computing is utilized. 

\begin{figure}
  \centering
\includegraphics[scale=0.65]{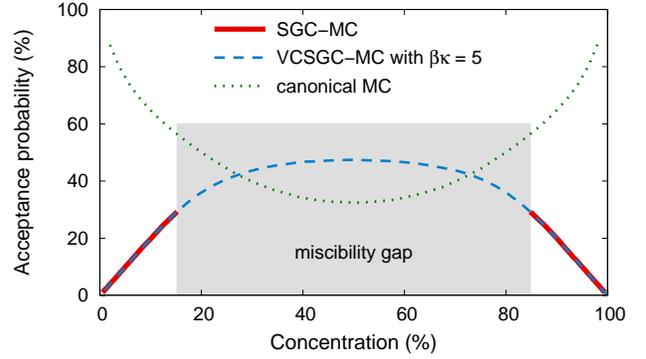}
  \caption{
    Comparison of the acceptance rate as a function of global concentration for the SGC, VC-SGC ($\kappa=5$), and canonical MC methods. At low concentrations the canonical MC method yields the highest acceptance rates while inside the miscibility gap the VC-SGC-MC achieves the best results.
  }
  \label{fig:vcsgc_accprob_conc}
\end{figure}

Thanks to its similarity with the SGC ensemble, it is straightforward to formulate a  MC algorithm for sampling the VC-SGC ensemble, where trial moves $\sigma^N\rightarrow\sigma_t^N$ comprise
\begin{enumerate}
\renewcommand{\theenumi}{\roman{enumi}}
\renewcommand{\labelenumi}{(\textit{\theenumi})}
\item selecting a particle at random,
\item flipping its spin,
\item computing the change in energy $\Delta U$ and concentration $\Delta c$ as well as
\begin{align}
  \tilde{c} =  \frac{\hat{c}(\sigma^N_t)^2 - \hat{c}(\sigma^N)^2}{2\Delta c} = 
  \frac{\hat{c}(\sigma^N_t) + \hat{c}(\sigma^N)}{2}.
  \label{eq:cbar}
\end{align}
\end{enumerate}
These trial moves are accepted with probability 
\begin{align}
  \mathcal{A}_{\VCSGC} =
  \min \left\{ 1,
  \exp\left[ -\beta \left(\Delta U + N\Delta c (\phi+2\kappa N \tilde{c})\right) \right]
  \right\}.
  \label{eq:trans_vcsgc}
\end{align}
Once again, this acceptance probability is designed to satisfy detailed balance. The force associated with a change in spin configuration receives contributions from both the change in the interatomic potential energy function $\Delta U$ as well 
as the external concentration dependent force $N\Delta c (\phi+2\kappa N \tilde{c})$. 
Hence, for a change in concentration, the usual interatomic forces are augmented with an additional external chemical driving force that at variance with the SGC ensemble is not  
a constant but varies linearly with concentration as $N\phi + 2\kappa N^2 c$. When ensemble-averaged, the equilibrium chemical driving force that corresponds to the chemical potential difference in physical experiments and the $\Delta\mu$ parameter in the SGC ensemble now becomes 
\begin{align}
  \Delta\mu = \phi + 2 \kappa N \left<\hat{c}\right>_{\VCSGC}.
  \label{eq:gforce}
\end{align}
This very important relation is derived in the appendix, see Eq.~\ref{eq:df1_vcsgc}. It connects the VC-SGC and the SGC ensembles and will be used extensively in the following to design and analyze Monte Carlo simulations of systems in which several phases coexist. 

We now apply the VC-SGC-MC method to study the model system described in \sect{sect:sgc}. Figure~\ref{fig:vcsgc_conc_mu_eps} illustrates the relation between the global concentration and the parameter ratio $\phi/\kappa$. It clearly demonstrates that using the VC-SGC-MC algorithm enables us to stabilize the system at arbitrary global concentrations in and outside the miscibility gap.

The dependence of the standard deviation of the concentration on the variance parameter $\kappa$ follows a power law [\fig{fig:vcsgc_stddev_eps}(a)], $\left<\Delta \hat{c}^2\right>_{\VCSGC} \propto 1/\sqrt{\kappa}$. The relation between the acceptance probability and $\kappa$, on the other hand, is linear with a negative slope [\fig{fig:vcsgc_accprob_eps}(b)]. Increasing $\kappa$ thus has two effects: It leads to a smaller standard deviation while simultaneously reducing the acceptance probability.

We can also compare the acceptance probability as obtained with the VC-SGC-MC method with the results for the SGC and canonical MC methods. As shown in \fig{fig:vcsgc_accprob_conc}, in the single-phase regions of the phase diagram the SGC and VC-SGC-MC methods coincide and produce comparably low acceptance rates, while the canonical MC method provides large acceptance rates. However, inside the miscibility gap, which is the region of interest when it comes to phase segregation, the VC-SGC method yields the best results.

We now study the functional dependence of the chemical driving force $\Delta\mu$ obtained from \Eq{eq:gforce} on the average concentration using the VC-SGC-MC method. The result is shown in \fig{fig:df1_vcsgc}(b) in comparison with the data obtained using the SGC-MC method. The VC-SGC-MC method produces a continuous relation between $\Delta\mu$ and $c$ throughout the entire concentration range. In the single-phase regions of the phase diagram the SGC and VC-SGC-MC results coincide. Inside the miscibility gap, where the SGC-MC fails, the VC-SGC-MC method reproduces the van-der-Waals loop associated with the formation of phase boundaries.\cite{Hil02} This is a very important result that can be used to extract interface free energies. \cite{SadErh12}

To summarize, the VC-SGC-MC method imposes a constraint on the variance of the concentration, and allows for equilibration at arbitrary global concentrations. Thereby, it merges the advantages of the SGC and the canonical MC algorithms. In the next section, we show that the VC-SGC-MC algorithm is also very well suited for parallelization enabling simulations of systems with many million particles.


\section{Parallelization strategies for large systems}
\label{sect:parallelization}

There are a multitude of problems involving precipitation, especially in the presence of structural defects such as dislocations, grain boundaries and surfaces, which require simulations of systems with hundreds of thousands or millions of particles. Efficient parallelization schemes with good scalability are a necessity in order to address these problems. Here, we focus on short-range interaction potentials as described e.g., by embedded-atom method, \cite{DawBas84} bond-order, \cite{Ter86} or Stillinger-Weber \cite{StiWeb85} type potentials. 

Monte Carlo simulations in the canonical ensemble do not lend themselves to efficient parallelization since trial moves in this scheme involve exchange of two particles that can be located on any two processors. Although it is possible to conceive elaborate distributed algorithms, it is difficult to implement a scheme that ensures unbiased sampling and still avoids spending a considerable fraction of simulation time on interprocessor communication. The SGC ensemble on the other hand can be parallelized easily but, as discussed in \sect{sect:sgc}, cannot be used to study precipitation. The purpose of this work is to develop a Monte Carlo technique that can both handle multiphase systems and can be parallelized easily and efficiently. In the following, we discuss parallelization strategies for the SGC as well as the VC-SGC ensembles and demonstrate their excellent scalability and efficiency.

\subsection{Domain decomposition for sampling trial moves}
\label{sect:decom}

Consider for simplicity a simulation box in the shape of a cube with linear dimension $L$. In systems with short-range interactions, the most common parallelization strategy is to subdivide the simulation box into a regular lattice of $N_{\CPU}$ equivalent cells $\{\mathcal{C}_i\}$ with linear dimension $L_c = L/N_c$, where $N_{\CPU} = N_c\times N_c \times N_c$. (
The generalization to non-cubic cells is straightforward). 

At every Monte Carlo step, a cubic domain $\mathcal{D}_i$ is chosen inside each cell $\mathcal{C}_i$ in such a way as to ensure that equivalent domains on different processors are {\em non-interacting}. This means that the total energy change $\Delta U$ associated with arbitrary spin flips inside the domains $\{\mathcal{D}_i\}$ can be written as the sum of the {\em independent} local energy changes $\Delta U_i$ on each processor, i. e. $\Delta U = \sum_{i=1}^{N_{\CPU}} \Delta U_i$. Note that all domains $\mathcal{D}_i$ are equivalent with linear dimension $L_D = L_c - R_c$, where $R_c$ is the effective interaction radius in the system. For pair interactions this radius equals the cutoff radius of the potential, while for three-body potentials it is usually twice the cutoff radius. 

\begin{figure}
  \centering
\includegraphics[width=0.88\columnwidth]{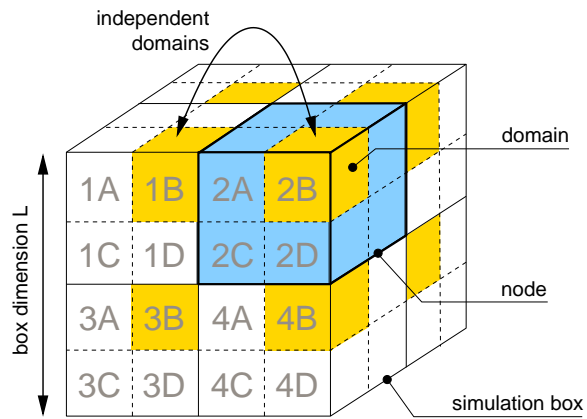}
  \caption{
    Spatial decomposition (solid lines) and subsequent division into octants (dashed lines) of a system with short-ranged interactions. Sets of octants with the same letter are independent of each other. One such set is marked in yellow.
  }
  \label{fig:decompos1}
\end{figure}

It is easy to see that for the above parallelization strategy to be possible the linear dimension $L_c$ must be larger than $R_c$. Let us first discuss the case when $L_c$ is exactly twice $R_c$. In this case the independent domains will have the linear dimension $L_D=R_c$. They constitute the eight non-overlapping octants of each cell $\mathcal{C}_i$ as depicted in \fig{fig:decompos1}. In this figure, all domains ``A'' are non-interacting and so are all domains ``B'' {\it etc}. At each Monte Carlo trial move, one of the eight octants is chosen at random. It is important that all cells $\mathcal{C}_i$ work on the same octant simultaneously since only in this way the trial moves on different processors are with certainty non-interacting.

\begin{figure*}
  \centering
\includegraphics[width=0.9\linewidth]{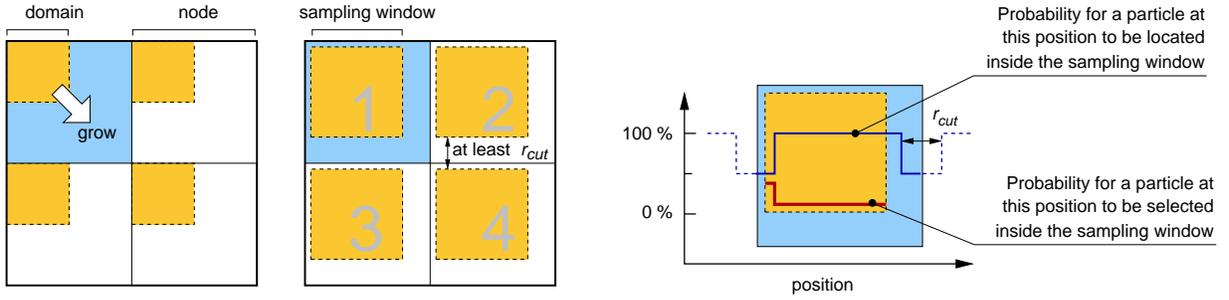}
  \caption{
    Schematic representation of an optimal spatial decomposition (compare \fig{fig:decompos1}). For a pair potential the domains have to be separated by at least one cutoff-distance while for potentials with three-body terms the separation has be to at least two cutoff-distances.
  }
  \label{fig:decompos2}
\end{figure*}

The above method of subdividing each cell $\mathcal{C}_i$ into eight non-overlapping octants also works when $L_C > 2R_c$ . However, bear in mind that confining the local trial moves to non-interacting domains produces weak spatial correlations that can slow down the approach to equilibrium, especially when phase segregation and growth of precipitates is expected. These spatial correlations are minimized if the total volume of the domains $\{\mathcal{D}_i\}$ is maximized. This can be achieved by growing each octant to a cube with linear dimension $L_D = L_c - R_c$. The eight distinct domains thus generated inside each cell $\mathcal{C}_i$ do overlap. This leads to the central region of $\mathcal{C}_i$ be covered by all eight $\mathcal{D}_i$. To ensure uniform sampling, the particles in the outer regions of the $\mathcal{C}_i$ cells must be selected with higher probability than those in the center. This can be achieved by assigning differential weights to the particles in the system depending on their position inside $\mathcal{C}_i$ (see the right panel of \fig{fig:decompos2}) prior to making trial moves. 

It is now straightforward to devise an efficient parallel Monte Carlo algorithm, where each trial move is composed of $N_{\CPU}$ local moves $\sigma^N_t\rightarrow\sigma^N_t+\Delta\sigma^N_t(i)$ carried out inside the domains $\{\mathcal{D}_i\}$ synchronously on all processors. To ensure uniform sampling, a trial move is constructed in two stages: ({\em i}) select one of the eight independent domains $\{\mathcal{D}_i\}$ at random and broadcast to all processors; message passing can be avoided by synchronizing the seed for the random number generator on all processors, and ({\em ii}) on each processor $i$, pick a particle at random inside the chosen domain and flip its spin. Different parts of the domain may be sampled with different weights. 
 
It is important to note that the composite trial move $\sigma^N\rightarrow\sigma^N+\sum_{i=1}^{N_{\CPU}} \Delta \sigma^N_t(i)$ constructed in this way will be rejected at a very high rate. In the following section, we describe how one can improve the above procedure in order to obtain reasonable acceptance probabilities for composite trial moves.

\subsection{Parallel Monte Carlo algorithms}

\subsubsection{Monte Carlo sampling of SGC ensemble}
\label{sect:sgc_parallelization}

In this section, we describe how one can devise parallel Monte Carlo simulations in the SGC-ensemble with composite trial moves constructed from trial moves simultaneously generated on all processors. The algorithm is as follows: 
({\em i}) On each processor $i$ make a local trial move $\Delta \sigma^N_t(i)$ according to one of the procedures described in section \ref{sect:decom}, 
({\em ii}) compute the local changes in energy $\Delta U_i$ and concentration $\Delta c_i$, and accept this move with probability 
\begin{align}
  \mathcal{A}^p_{\SGC}(i) =
  \min \left\{ 1,
  \exp\left[ -\beta (\Delta U_i + \Delta\mu N\Delta c_i) \right]
  \right\},
  \label{eq:trans_sgc_par}
\end{align}
otherwise set $\Delta \sigma^N_t(i) = 0$. The global composite trial move is now $\sigma^N\rightarrow\sigma^N+\sum_{i=1}^{N_{\CPU}} \Delta \sigma^N_t(i)$.  Thanks to the independence of the domains $\mathcal{D}_i$, the transition probability for this move is proportional to $\prod_{i=1}^{N_{\CPU}}\mathcal{A}^p_{\SGC}(i)$ and satisfies detailed balance. 


\subsubsection{Monte Carlo sampling of VC-SGC ensemble}
\label{sect:vcsgc_parallelization}

The similarity of the SGC and VC-SGC ensembles discussed in \sect{sect:vcsgc} suggests that parallelization strategies might be similar as well. A closer inspection, however, reveals that for a composite trial move $\sigma^N\rightarrow\sigma^N_t$, where $\sigma^N_t=\sigma^N+\sum_{i=1}^{N_{\CPU}}\Delta \sigma^N_t(i)$, we have
\begin{align}
  \hat{c}\left(\sigma^N_t\right)^2 - 
  \hat{c}\left(\sigma^N\right)^2 \neq
  \sum_{i=1}^{N_{\CPU}}\hat{c}\left(\sigma^N+\Delta \sigma^N_t(i)\right)^2.
\end{align}
This implies that there is a coupling between the domains $\mathcal{D}_i$, and as a result the simple method outlined in the previous section for the SGC ensemble cannot be directly applied to the parallel sampling of the VC-SGC ensemble. To resolve this issue, we first modify the acceptance probability distribution \Eq{eq:trans_vcsgc} for the serial sampling of the VC-SGC ensemble as follows
\begin{align}
  \mathcal{A}_{\VCSGC} =
  &\min \left\{ 1,
  \exp\left[ -\beta \left(\Delta U
    + N\Delta c (\phi+2\kappa N c_0)\right) \right]
  \right\} \nonumber
  \\
  &\times
  \min \left\{ 1,
  \exp\left[ -\beta \kappa N^2 \Delta c (\tilde{c}-c_0)\right]
  \right\} 
\label{eq:mod}
\end{align}
where $\tilde{c}$ was defined in \Eq{eq:cbar}. It is easy to verify that the acceptance probability distribution in \Eq{eq:mod} satisfies detailed balance. The parameter $c_0$ introduced in \Eq{eq:mod} can change the acceptance probability and thus the approach to equilibrium but it does not affect the final result. An optimal choice is
\begin{align}
\label{eq:c0}
  c_0 = \left<\hat{c}\right>_{\VCSGC}. 
\end{align}
In practice, the simulations are performed with $c_0$ chosen to be the best guess for the average concentration. In \sect{sect:par_efficiency}, we will explicitly demonstrate the correlation between $c_0$, $\kappa$, $\phi$, and discuss acceptance rates for the simple Ising model introduced earlier.

We can now formulate a parallel Monte Carlo algorithm in the VC-SGC ensemble with composite trial moves comprising $N_{\CPU}$ synchronous local moves $\sigma^N\rightarrow\sigma^N+\sum_{i=1}^{N_{\text{CPU}}}\Delta\sigma^N_t(i)$:
({\em i}) on each processor $i$ make a local trial move as detailed in the \sect{sect:decom},
({\em ii}) compute the local changes in energy $\Delta U_i$ and concentration $\Delta c_i$, and accept this move with probability 
\begin{align}
  \mathcal{A}^{p,\text{loc}}_{\VCSGC}(i) =
  \min \Big\{ 1,
  \exp \big[ -\beta (
    \Delta U_i
    + N\Delta c_i \underbrace{(\phi + 2\kappa N c_0)}_{\displaystyle{=\Delta\mu_0}}
    ) \big]
  \Big\},
  \label{eq:trans_vcsgc_par1}
\end{align}
otherwise set $\Delta \sigma^N_t(i)=0$. Following \Eq{eq:mod}, the global trial move $\sigma^N\rightarrow\sigma^N+\sum_{i=1}^{N_{\CPU}}\Delta \sigma^N_t(i)$ may be accepted with probability
\begin{align}
  \mathcal{A}^{p,\text{glob}}_{\VCSGC}
  &= 
  \min \left\{ 1,
  \exp\left[ -2 \beta \kappa N^2 \sum_i \Delta c_i (\tilde{c}_i-c_0)\right]
  \right\}
  \nonumber \\
  &= \min \left\{ 1,
  \exp\left[ -\beta \kappa N^2 \Delta c_\text{tot} \left( 
  \Delta c_\text{tot} - 2(\hat{c}(\sigma^N)-c_0) \right) \right]
  \right\},
  \label{eq:trans_vcsgc_par2}
\end{align}
where $\Delta c_\text{tot}=\sum_{i=1}^{N_{\CPU}} \Delta c_i$ is the total change in concentration due to the composite trial move. This quantity can be efficiently computed using for example the message passing interface \cite{mpi} \texttt{allgather} command.


\subsection{Efficiency of the parallel VC-SGC-MC method}
\label{sect:par_efficiency}

In arriving at \Eq{eq:trans_vcsgc_par1}, we have introduced the parameter $c_0$ and the abbreviation $\Delta\mu_0=\phi + 2 \kappa N c_0$. Together with $\kappa$ these parameters determine the average and the variance of the concentration. In this section, we will demonstrate the correlation between these parameters using the simple Ising model described in \sect{sect:sgc}.

The derivation of the transition matrix for the parallel VC-SGC-MC method in the previous section revealed a close resemblance with the parallel SGC-MC method. In particular, the acceptance probabilities $\mathcal{A}^p_{\SGC}(i)$ and $\mathcal{A}^{p,\text{loc}}_{\VCSGC}(i)$ in Equations~\eq{eq:trans_sgc_par} and \eq{eq:trans_vcsgc_par1} become identical if $\Delta\mu_0=\Delta\mu$. This of course requires $c_0$ to be chosen according to the optimality condition \Eq{eq:c0}. This insight greatly simplifies the choice of parameters for the parallel VC-SGC-MC method.

\begin{figure}
  \centering
  \includegraphics[scale=0.65]{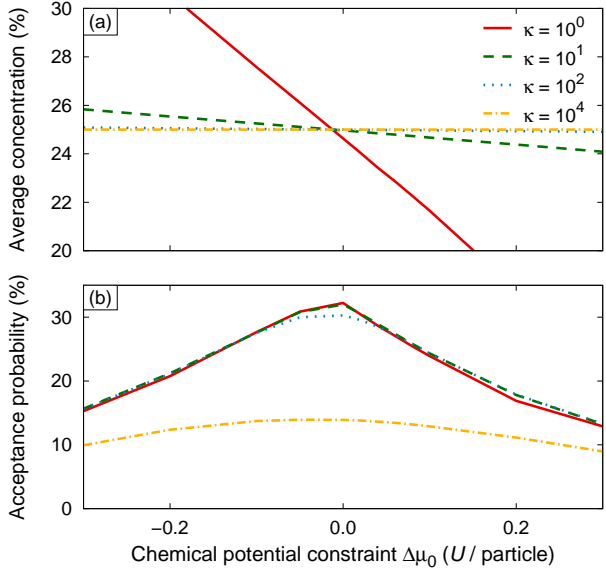}
  \caption{
    (a) Average concentration and (b) acceptance probability obtained from parallel VC-SGC-MC simulations using different combinations of the parameters $\Delta\mu_0$ and $\kappa$ for $c_0=0.25$ in \Eq{eq:trans_vcsgc_par2}. 
  }
  \label{fig:conc_dmu_vcsgc}
  \label{fig:accprob_dmu_vcsgc}
\end{figure}

In \fig{fig:conc_dmu_vcsgc}(a), we show the average concentration obtained in simulations with different values of $\Delta\mu_0$ and $\kappa$, for a fixed target concentration of $c_0=0.25$ located inside the miscibility gap. All simulations were carried out using 64 CPUs, a $4\times 4\times 4$ decomposition, and a BCC lattice with $40\times 40\times 40$ conventional unit cells. The number of particles per processor is thus equal to the number of particles in the serial VC-SGC-MC simulations discussed in \sect{sect:vcsgc}.

For small values of $\kappa$ the average concentration varies strongly with $\Delta\mu_0$. As $\kappa$ is increased, the total concentration is confined to small variations about $c_0$ and the average concentration becomes less sensitive to the choice of $\Delta\mu_0$. Comparison with \fig{fig:df1_vcsgc}(b), where the chemical driving force is shown as a function of average concentration $\left<\hat{c}\right>$, reveals that $\left<\hat{c}\right>$ equals $c_0$ exactly when $\Delta\mu=\Delta\mu_0$. This confirms \Eq{eq:c0} and validates the underlying connection between the SGC and VC-SGC-MC methods. 

\begin{figure}
  \centering
\includegraphics[scale=0.65]{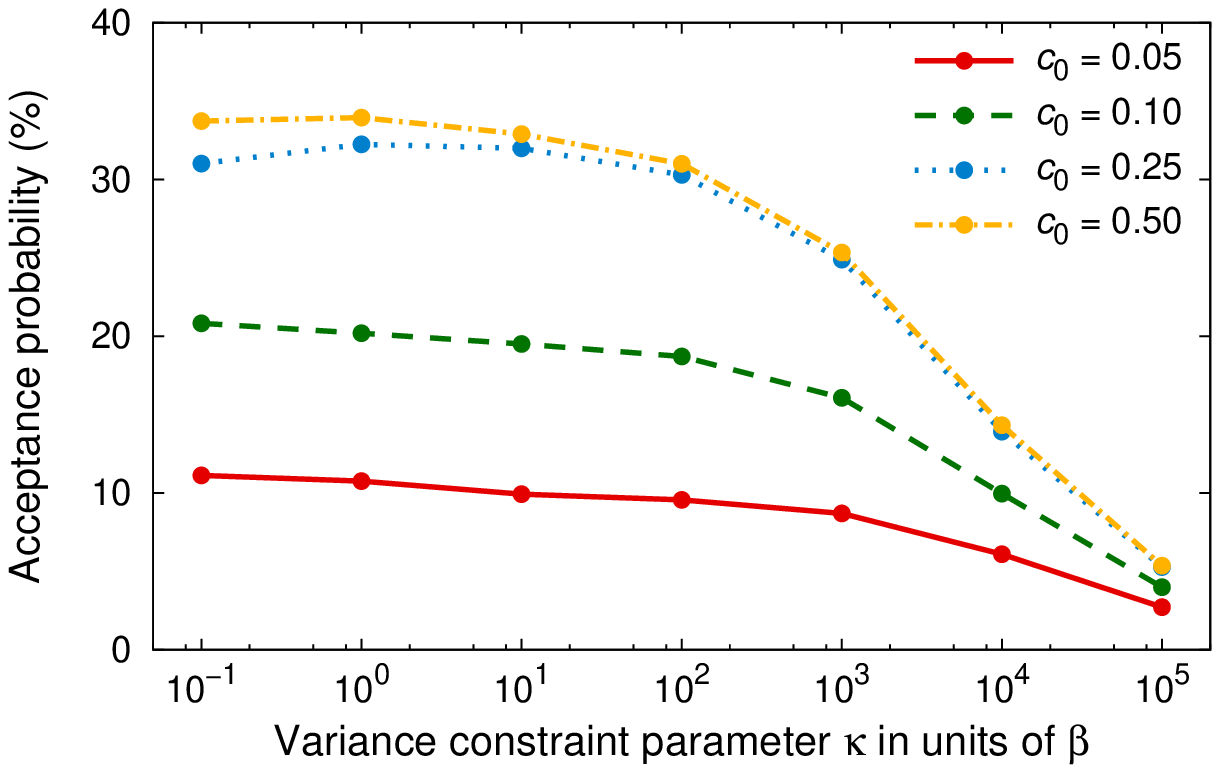}
  \caption{
    Acceptance probability as a function of variance constraint parameter $\kappa$ for different target concentrations $c_0$ and optimal values for $\Delta\mu_0$ as obtained with the parallel VC-SGC-MC algorithm.
  }
  \label{fig:accprob_eps_par}
\end{figure}

While for sufficiently large values of $\kappa$ the parameter $\Delta\mu_0$ does not affect the average concentration, it does have a significant impact on the acceptance probability as illustrated in \fig{fig:accprob_dmu_vcsgc}(b). For a given value of $\kappa$ the acceptance probability becomes maximal if $\Delta\mu=\Delta\mu_0$, which again confirms the optimality condition \Eq{eq:c0}. Similar to the case of the serial VC-SGC-MC algorithm [compare   \fig{fig:vcsgc_accprob_eps}(b)], we also find that for a fixed value of $\Delta\mu_0$, the acceptance probability decreases with increasing $\kappa$ as shown explicitly in \fig{fig:accprob_eps_par}. It is however remarkable that over a rather wide range the value of $\kappa$ does not have a significant negative impact on the acceptance probability.

Now that we have understood the interplay between the parameters $\Delta\mu_0$, $\kappa$, and $c_0$, we can formulate an optimal strategy for choosing their values:
\begin{enumerate}
\item[({\em i})]
  Determine the chemical driving force $\Delta\mu_S$ in the vicinity of the two-phase region using the SGC-MC method. This requires simulations involving only small system sizes since we are only interested in single-phase equilibria [compare \fig{fig:df1_vcsgc}(b)].
\item[({\em ii})]
  Choose a value of $\kappa$ based on the desired standard deviation of the concentration (compare \fig{fig:vcsgc_stddev_eps}).
\item[({\em iii})]
  Set $\Delta\mu_0=\Delta\mu_S$ and $c_0$ to the desired concentration inside the miscibility gap. In this way the parameter $\phi = \Delta\mu_0 - 2\kappa N c_0$ is determined as well. For all subsequent simulations inside the miscibility gap $\Delta\mu_0$ can be held fix and only $c_0$ is tuned to obtain the desired concentration.
\end{enumerate}
  
From \fig{fig:accprob_eps_par} one observes that at an average concentration of 50\%\ the parallel VC-SGC-MC algorithm achieves a maximal acceptance ratio of about 34\%\ which compares favorably with a maximum value of about 47\%\ for the serial VC-SGC-MC method (see \fig{fig:vcsgc_accprob_eps}).

\begin{figure}
  \centering
\includegraphics[scale=0.65]{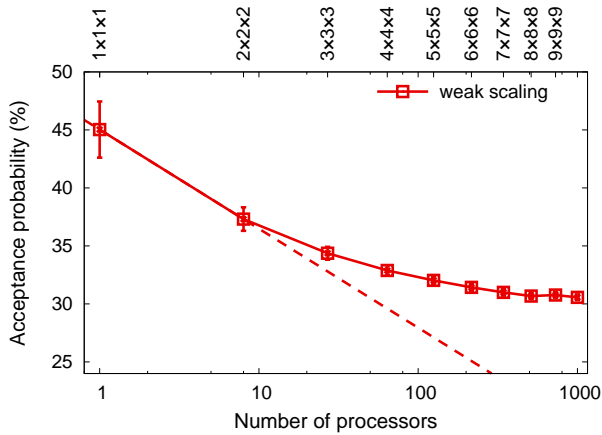}
  \caption{
    Acceptance probability as a function of the number of processors both in the weak scaling limit using $\Delta\mu_0=0$, $\kappa=10$, and $c_0=0.5$. The dashed line represents logarithmic scaling.
  }
  \label{fig:scaling}
\end{figure}

To investigate the performance of the parallel VC-SGC-MC algorithm in the weak scaling limit, a series of simulations with an increasing number of processors was carried out in which the number of particles per processor was kept constant (2,000 particles, $10\times 10\times 10$ conventional unit cells) while the total system size was increased along with the number of processors. The results of the scaling analysis are summarized in \fig{fig:scaling}. As can be seen by comparison with the dashed line, in the weak scaling limit, the  acceptance probability scales better than logarithmically with the number of processors. These results provide clear evidence that the VC-SGC-MC algorithm is ideally suited for simulations of very large systems.

The good scalability of the algorithm can be rationalized as follows: In the first part of a VC-SGC-MC trial step, a composite move is constructed that in the second part is accepted or rejected as a whole. The collective acceptance/rejection of a large number of individual moves could suggest that the acceptance probability for the second rejection decreases rapidly with the number of individual moves and thus the number of processors. The first acceptance/rejection, however, ensures that the combination of the individual moves form a cluster move that is already ``optimized'' and therefore has a relatively low probability to be rejected in the second part of the VC-SGC-MC trial move.


\section{Application to realistic alloys}
\label{sect:applications}

\subsection{Sampling structural relaxation and vibrations}

\begin{figure*}
  \centering
\includegraphics[width=0.9\linewidth]{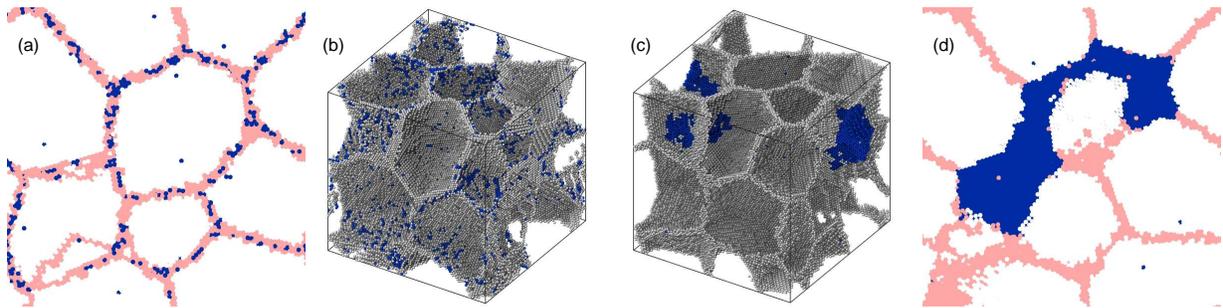}
  \caption{
    (Color online)
    Representative snapshots obtained after full equilibration in simulations using the (a,b) LF potential and (c,d) the PM potential. Coloring according to common-neighbor analysis. (a,d) BCC Fe atoms, Fe and Cu grainboundary atoms are shown in white, pink, and blue, respectively. (b,c) Fe and Cu grainboundary atoms are shown in gray and blue, respectively.
  }
  \label{fig:FeCu_nanocryst}
\end{figure*}

In the previous sections, we have discussed in detail the development of an efficient parallel MC algorithm for studying systems with millions of particles at arbitrary global concentrations. For the purpose of demonstration, we considered a simple lattice model. In many practical applications, however, the configuration space includes continuous particle coordinates leading to structural relaxations and thermal vibrations.

As shown in \sect{sect:canonical}, structural and chemical degrees of freedom can be separated readily in the partition function. This allows us to sample the corresponding integrals with different techniques. A straightforward approach is to combine transmutation and displacement MC trial moves. In practice, this algorithm, however, often converges poorly especially when structural relaxations are involved. As indicated after \Eq{eq:trans_canonical}, a much more efficient way to sample displacements is obtained by combining transmutation Monte Carlo moves with molecular dynamics simulations. In practice, one carries out a MD simulation that is interrupted every $n$-th MD step to execute $m$ MC trial moves. While optimal sampling is obtained if $n=m=1$ [compare comment after \Eq{eq:trans_canonical}], for computational efficiency it is beneficial to choose larger values. This does not affect sampling as long as the total number of MD/MC cycles is sufficiently large, i. e. $n$ is much smaller than the total number of MD steps.

We have applied the hybrid MC/MD approach for modeling chemical ordering and/or precipitation in several metallic alloys in the vicinity of heterogeneities such as dislocations, grain boundaries and surfaces. In the next section, we consider the precipitation of Cu in Fe-rich Fe--Cu nanocrystals as an illustration for the type of problems that can be studied using our algorithm. Other examples include the study of grain boundary pinning in Cu due to Fe impurities \cite{ZepGilSad05}, structural phase transformations of Cu precipitates in BCC iron \cite{ErhMarSad12}, short-range order in Fe--Cr alloys, \cite{ErhCarCar08} and the properties of helium bubbles in Fe and Fe--Cr alloys \cite{CarHetStu11}.

can be found in Ref.~\onlinecite{ErhCarCar08}, where we used a preliminary version of the present algorithm to study short-range order in Fe--Cr alloys as a function of temperature and composition.

\subsection{Cu precipitation in Fe nanocrystals}

We will now concern ourselves with VC-SGC-MC/MD simulations of Cu-precipitation in dilute nanocrystalline ferritic Fe--Cu alloys. The very small solubility of Cu in Fe (0.07\%\ at 700\,K) gives rise to a very strong driving force for precipitation. The different crystal structures of Fe (body-centered cubic, BCC) and Cu (face-centered cubic, FCC) as well as the mechanical instability of bulk BCC-Cu, imply that as Cu precipitates grow structural phase transformations occur. This realization in conjunction with the technological importance of Fe--Cu alloys has lead to a considerable amount of research in this field (see e.g., Refs.~\onlinecite{OthJenSmi91, BlaAck01}). Here, we compare the precipitation of Cu in dilute nanocrystalline Fe--Cu alloys simulated using two different interatomic potential models.

First, a nanocrystalline BCC-Fe sample with dimensions of 18.8\,nm in all Cartesian directions was created as follows. Eleven randomly oriented BCC seeds (average grain diameter 4\,nm) were distributed evenly in the simulation cell and nanocrystallites were constructed by filling the Voronoi volumes around each seed. The resulting grain boundary structure was relaxed using conjugate gradient minimization and subsequently equilibrated at a temperature of 700\,K for 500\,ps using MD simulations. The final sample contained 548,565 atoms.

VC-SGC-MC/MD simulations were performed at 700\,K using $\Delta\mu_0=-0.60\,\eV$ in \Eq{eq:trans_vcsgc_par1}, $\kappa=1000$ in \Eq{eq:trans_vcsgc_par2}, and a target concentration $c_0=4\%$. One MC cycle (equivalent to $N_{at}$ MC trial moves where $N_{at}$ is the number of atoms) was carried out per 20 MD steps. The equations of motion were integrated for 1,200,000 MD steps (including 60,000 MC cycles) using a time step of 2.5\,fs. Temperature and pressure were maintained using the Nos\'e-Hoover thermostat and barostat, respectively.

Interatomic interactions were modeled using both the Fe--Cu potential by Ludwig \etal\ \cite{LudFarPed98} (LF) and the potential by Pasianot and Malerba \cite{PasMal07} (PM). The LF potential is based on the Fe potential by Simonelli \etal\ \cite{SimPasSav93} and the Cu potential by Voter, \cite{Vot93} while the PM potential employs the Fe potential by Mendelev \etal\ \cite{MenHanSro03} and the Cu potential by Mishin \etal\ \cite{MisMehPap01}. Both potentials give solubilities at 700\,K that are very close to the experimental value (LF: 0.15\%, PM: 0.07\%, experiment: approximately 0.07\%), and thus the target concentration of 4\%\ is far beyond the solubility limits for either potential.

Figure~\ref{fig:FeCu_nanocryst} summarizes the key results of our analysis. As expected, both potentials show a very small number of Cu atoms in the center of the grains. As the total Cu concentration of about 4\%\ is far larger than the bulk solubility this implies that excess copper is located in grain boundaries. While the two potentials agree with regard to the latter trend, they yield very different results when it comes to the {\em distribution} of the Cu in the grain boundaries. Whereas the LF potential predicts a homogeneous distribution with little spatial correlation between the Cu atoms [see \fig{fig:FeCu_nanocryst}(a,b)], the PM potential yields contiguous Cu precipitates that are agglomerated along only a few neighboring grain boundaries. While this result showcases the kind of insight that can be gained using the VC-SGC-MC/MD hybrid simulation technique, it also demonstrate that further work in the area of potential development and verification is needed before a reliable study of Cu precipitation at grainboundaries in Fe can be conducted.


\section{Conclusions}
\label{sect:conclusions}

In the present paper, we have developed a hybrid molecular dynamics/Monte Carlo (MD/MC) algorithm which is ideally suited for simulating multicomponent systems using samples with millions of particles in both single and multiphase regions of the phase diagram. The most important component is an efficient and scalable transmutation MC method that samples the variance-constrained semi-grandcanonical ensemble. The VC-SGC-MC algorithm can be used to stabilize multiphase equilibria and therefore allows to study precipitation and phase segregation. Since it features a better-than-logarithmic scaling of the acceptance probability with the number of processors, the method is ideally suited for studying very large systems containing several million particles. Finally, by combining the VC-SGC-MC method with molecular dynamics, one obtains a very powerful hybrid scheme that takes into account chemical mixing and precipitation, structural relaxations as well as thermal vibrations.

We have applied the algorithm developed in this work to study the precipitation of Cu in nanocrystalline Fe using two different interatomic potentials. While both potentials predict excess Cu to be located in the grain boundaries, they yield very different results for the distribution of impurity atoms in the grain boundaries. Further work in potential development and verification is required in order to obtain interatomic potential models that provide reliable predictions for element distribution near inhomogeneities such as dislocations, grain boundaries, and surfaces.

The hybrid MC/MC algorithm described in this paper has already been applied to study for example grain boundary pinning in Cu due to Fe impurities \cite{ZepGilSad05}, structural phase transformations of Cu precipitates in BCC Fe \cite{ErhMarSad12}, short-range order in Fe--Cr alloys, \cite{ErhCarCar08} and the properties of helium bubbles in Fe and Fe--Cr alloys \cite{CarHetStu11}. The relation to free energy integration that is implicit to Eq.~\eq{eq:df1_vcsgc} has furthermore been utilized in Ref.~\cite{SadErh12} to obtain the temperature and orientation dependence of free interface energies in Fe--Cr alloys.

The algorithms developed in the present work have been implemented in the massively parallel MD code \textsc{lammps}. \cite{Pli95} The source code is available from the authors.

\begin{acknowledgments}
  Lawrence Livermore National Laboratory is operated by Lawrence Livermore National Sec\-urity, LLC, for the U.S. DOE-NNSA under Contract DE-AC52-07NA27344. Com\-puter time allocations by NERSC at Lawrence Berkeley National Laboratory and the Swedish National Infrastructure for Computing are gratefully acknowledged. One of us (PE) has been partly supported by a grant from the Swedish Research Council.
\end{acknowledgments}


\appendix*
\section {Derivation of the VC-SGC ensemble}

In this appendix, we derive the VC-SGC ensemble for the binary system discussed in section~\ref{sect:chemical_mixing}. Consider a system of $N$ particles confined in a box of volume $V$, where each particle carries a spin of value 0 or 1. Since the VC-SGC ensemble only manipulates the chemical degrees of freedom we consider for simplicity a system frozen onto a lattice of an arbitrary configuration $\vec{x}^{3N}$. The phase space $\Omega$ of this system consists of the set of $\rho=2^N$ different spin configurations $\{\sigma^N\}$. To simplify the notation below, we enumerate the $\rho$ configurations in $\Omega$: $\{1,2,\cdots,\rho\}$, and thus any spin configuration $\sigma^N$ can be uniquely identified by its index number. 

Let $\Sigma$ be the set of $M$ representative configurations in $\Omega$ and denote by $n_{\alpha}$ the number of times the $\alpha$-th state appears in $\Sigma$. We can uniquely define $\Sigma$ by the set of numbers $\{n_1,n_2,...,n_{\rho}\}$. The sum of the occupation numbers $n_{\alpha}$ are constrained according to
\begin{align}
  M &= \sum_{\alpha=1}^{\rho} n_{\alpha} .
  \label{eq:M}
\end{align}
We now introduce three more constraints for ({\it i}) the average energy $\overline{U}$, ({\it ii}) the average concentration of spin zero particles $\overline{c}$, and ({\it iii}) the square of the concentration of spin zero particles $\overline{v}^2$. These constraints can be expressed as:
\begin{align*}
  \overline{U}
  &= \frac{1}{M}\sum_{\alpha=1}^{\rho} n_{\alpha}~U(\alpha),
  \\
  \overline{c}
  &= \frac{1}{M}\sum_{\alpha=1}^{\rho} n_{\alpha}~\hat{c}(\alpha),
  \\
  \overline{v}^2
  &= \frac{1}{M}\sum_{\alpha=1}^{\rho} n_{\alpha}~\hat{c}(\alpha)^2
  .
\end{align*}
Above, we have denoted the potential energy for the state $\alpha$ by $U(\alpha)$ and its concentration by $\hat{c}(\alpha)$. For any given set $\Sigma=\{n_{\alpha}\}$, there are multiple ways of choosing its elements from $\Omega$. We use this to define the multiplicity $\eta$ of a set $\Sigma$:
\begin{align*}
  \eta &= \frac{M!}{\prod_{\alpha=1}^{\rho} n_{\alpha}!} .
\end{align*}
The relative probability of two sets with the same average energy $\overline{U}$ is now determined by the ratio of their multiplicities. In the thermodynamic limit, i.e. large $N$ and large $M$, the most probable $\Sigma$, i.e. the set with the largest multiplicity, will correspond to the equilibrium probability distribution in $\Omega$. Under the above constraints, the most probable distribution of $\{n_{\alpha}\}$ is determined by minimizing the functional $Q\left(\{n_{\alpha}\};\mu,\beta,\tilde{\phi},\tilde{\kappa}\right)$:
\begin{align}
  Q &=
  -\ln \eta
  - \mu \left(\sum_{\alpha=1}^{\rho} n_{\alpha} - M\right)
  \nonumber
  \\
  & \quad - \beta \left(\sum_{\alpha=1}^{\rho}
  n_{\alpha} E(\alpha) - M \overline{U}\right)
  \nonumber
  \\
  & \quad - \tilde{\phi} \left(\sum_{\alpha=1}^{\rho}
  n_{\alpha} \hat{c}(\alpha) - M \overline{c}\right)
  \nonumber
  \\
  & \quad - \tilde{\kappa} \left(\sum_{\alpha=1}^{\rho}
  n_{\alpha} \hat{c}(\alpha)^2 - M \overline{v}^2\right)
  \label{eq:Q} .
\end{align}
Above, $\mu$, $\beta$, $\tilde{\phi}$, and $\tilde{\kappa}$, are Lagrange multipliers that are introduced as independent variables to facilitate the constrained minimization of the functional $Q$ with respect to the occupation numbers $\{n_{\alpha}\}$. At its minimum, the derivative of the functional $Q$ with respect to the independent variables is set to zero. Setting $\partial Q/\partial n_{\alpha}$ to zero determines their equilibrium distribution:
\begin{align*}
  n_{\alpha} &=
  \exp\left[
    - \mu -\beta U(\alpha)
    - \tilde{\phi} \hat{c}(\alpha)
    - \tilde{\kappa} \hat{c}(\alpha)^2
 \right] .
\end{align*}
Using this result in \eq{eq:M} we obtain an explicit expression for the chemical potential $\mu$
\begin{align}
\label{eq:expmu}
  \exp(\mu) = \frac{1}{M}\sum_{\alpha=1}^{\rho}
  \exp\left[
    - \beta U(\alpha) - \tilde{\phi} \hat{c}(\alpha) - \tilde{\kappa} \hat{c}(\alpha)^2
    \right] .
\end{align}
Now it is possible to define the equilibrium probability of any state $\alpha$ in $\Omega$ as
\begin{align}
  \pi_{\VCSGC}({\alpha})
  &=
  \mathcal{Z}_{\VCSGC}^{-1} \exp\left[
    - \beta \left(U(\alpha) + N\hat{c}(\alpha)(\phi +\kappa N\hat{c}(\alpha))\right)
    \right] , 
  \label{eq:vprob}
\end{align}
where $\mathcal{Z}_{\VCSGC} = M e^{\mu}$, and we have introduced the definitions
\begin{align}
  \tilde{\phi} &= N\beta\phi\label{eq:tilde}\\
  \tilde{\kappa} &= N^2\beta\kappa\nonumber, 
\end{align}
in order to reproduce the equilibrium probability distribution of the VC-SGC ensemble Eq.~\eq{eq:pi_vcsgc}. Let us now define the phase space $\Omega_c$  of configurations with a fixed concentration $c$. The canonical free energy $F_{\C}(c)$ for this set can be defined as follows
\begin{align}
  \label{eq:canon}
  \exp\left[ - \beta F_{\C}(c) \right]
  &= \sum_{\alpha\in\Omega_c} \exp\left[-\beta E(\alpha) \right].
\end{align}
In this way the partition function Eq.~\eq{eq:expmu} can be rewritten as 
\begin{align}
  \mathcal{Z}_{\VCSGC} =  \int_0^1 \exp\left[
    - \beta \left(F_{\C}(c) + N c(\phi +\kappa Nc)\right)
\right] dc. 
\label{eq:cond0}
\end{align}

Setting $\partial Q/\partial \tilde{\phi}$ and $\partial Q/\partial \tilde{\kappa}$ in \eq{eq:Q} to zero and using the definitions \eq{eq:tilde} and \eq{eq:canon} provides for a system of two equations to determine the two  unknowns $\phi$ and $\kappa$
\begin{align}
\label{eq:cond1}
  \overline{c}
  &= \mathcal{Z}^{-1}
  \int_0^1 c \exp\left[
    - \beta \left(F_{\C}(c) + N c(\phi +\kappa Nc)\right)
    \right] dc
  \\
\label{eq:cond2}
  \overline{v}^2
  &= \mathcal{Z}^{-1}
  \int_0^1 c^2 \exp\left[ 
- \beta \left(F_{\C}(c) + N c(\phi +\kappa Nc)\right).
  \right] dc .
\end{align}

In solving the above equations, we assume that $\overline{v}$ is chosen such that it is much smaller than $\overline{c}$ and $1-\overline{c}$. Then it is possible to represent $F_\C(c)$ by its Taylor expansion to second order around $\overline{c}$:
\begin{align*}
  F_{\C}(c)\
  &= F_{\C}(\overline{c})
  + \left.\frac{\partial F_{\C}}{\partial c}\right|_{\overline{c}} (c-\overline{c})
  + \frac{1}{2} \left.\frac{\partial^2 F_{\C}}{
    \partial c^2}\right|_{\overline{c}} (c-\overline{c})^2, 
\end{align*}
and replace the integrals in Eqs.~(\ref{eq:cond0}--\ref{eq:cond2}) above with indefinite Gaussian integrals
\begin{align}
  1 &=
  \widetilde{\mathcal{Z}}_{\VCSGC}^{-1} \int_{-\infty}^{\infty}
  \exp\left[ - A (c-\overline{c}) - B (c-\overline{c})^2 \right] dc
  \nonumber
  \\
  \overline{c} &=
  \widetilde{\mathcal{Z}}_{\VCSGC}^{-1} \int_{-\infty}^{\infty}
  c \exp\left[ - A (c-\overline{c}) - B (c-\overline{c})^2 \right] dc
  \label{eq:eqset}
  \\
  \overline{v}^2 &=
  \widetilde{\mathcal{Z}}_{\VCSGC}^{-1} \int_{-\infty}^{\infty}
  c^2 \exp\left[ - A (c-\overline{c}) - B (c-\overline{c})^2 \right] dc
  \nonumber
\end{align}
where 
\begin{align*}
  A &= \beta \left[ \left.\frac{\partial F_{\C}}{\partial c}\right|_{\overline{c}} + 
     N\left(\phi +2\kappa N\overline{c}\right)
     \right]
  \\
  B &= \beta \left[ \frac{1}{2} \left.\frac{\partial^2 F_{\C}}{
    \partial c^2}\right|_{\overline{c}}+N^2 \kappa \right]
  \\
  \widetilde{\mathcal{Z}}_{\VCSGC} &= \mathcal{Z}_{\VCSGC}
  \exp\left[ \beta \left( F_{\C}(\overline{c}) + N\overline{c}\left(\phi + \kappa N\overline{c} \right) \right) \right] .
\end{align*}

It is now easy to see that the system of equations~\eq{eq:eqset}
can be satisfied when
\begin{align*}
  A = 0
  \quad \text{and} \quad
  B = \frac{1}{2 (\overline{v}^2-\overline{c}^2)}
\end{align*}

Hence within the VC-SGC ensemble, the thermodynamic forces ($\phi$ and
$\kappa$) that give rise to a given average concentration $\overline{c}$
and its standard deviation $s_0 = \sqrt{\overline{v}^2-\overline{c}^2}$, are related to
the derivatives of the Helmholtz free energy at $\overline{c}$ as follows
\begin{align*}
  N\phi &=
  \left.\frac{\partial^2 F}{ \partial c^2}\right|_{\overline{c}}
  - \left.\frac{\partial F}{
    \partial c}\right|_{\overline{c}} - \frac{\overline{c}}{\beta s_0^2}
  \\
  N^2\kappa &=
  \frac{1}{2}\left(\frac{1}{\beta s_0^2}
  - \left.\frac{\partial^2 F}{\partial c^2}\right|_{\overline{c}}
  \right) .
\end{align*}

The first derivative of the free energy with respect to the
concentration of one species, i.e. the difference in chemical potential between the two species $\Delta \mu$, can therefore be
calculated from the average concentration according to
\begin{align}
  \Delta \mu \equiv -\frac{1}{N}\left.\frac{\partial F_{\C}}{\partial c}\right|_{\overline{c}}
  &=
  \phi + 2 \kappa N \overline{c}.
  \label{eq:df1_vcsgc}
\end{align}
We have thus arrived at the important relation Eq.\eq{eq:gforce}, which is used extensively 
in this paper. In the same way, 
a similar relation can also be obtained between the second derivative and
the variance of the concentration which reads
\begin{align}
  -\left.\frac{\partial^2 F_{\C}}{\partial c^2}\right|_{\overline{c}}
  &=
  2N^2\kappa - \frac{1}{\beta s_0^2} .
  \label{eq:df2_vcsgc}
\end{align}


%

\end{document}